\documentclass[runningheads]{llncs}

\usepackage[latin1]{inputenc}
\usepackage{amsmath,amssymb}
\usepackage{amssymb}
\usepackage{graphicx}
\usepackage{xcolor,soul}
\usepackage{adjustbox}
\usepackage{listings}

\usepackage[final=true]{hyperref}

\usepackage{soul, color}

\lstdefinelanguage{SPARQL}{
	keywords={@prefix, @base, PREFIX, BASE, LIMIT, SELECT, SERVICE, GRAPH, CONSTRUCT, WHERE, FILTER, GENERATE, BIND, SOURCE, ITERATOR, AS},
	keywordstyle=\color{blue}\bfseries,
	ndkeywords={a, class, predicate, subjectMap, template, dul, schema, rr, rml, xsd, ql, ex, objectMap, predicateObjectMap},
	ndkeywordstyle=\color{darkgray}\bfseries,
	identifierstyle=\color{black},
	sensitive=false,
	comment=[s]{<}{>},
	commentstyle=\color{purple}\ttfamily,
	stringstyle=\color{red}\ttfamily,
	morestring=[b]',
	morestring=[b]"
}

\begin{document}

%

\title{Towards Integrated and Open COVID-19 Data\thanks{The research work was supported by the Hellenic Foundation for Research and Innovation (H.F.R.I.) under the ``First Call for H.F.R.I. Research Projects to support Faculty members and Researchers and
the procurement of high-cost research equipment grant'' (Project Number: HFRI-FM17-81).
}}
%
%
\author{Georgios M. Santipantakis \and George A. Vouros \and
	Christos Doulkeridis}
\authorrunning{F. Author et al.}
%
\institute{University of Piraeus, Greece
\email{\{gsant,cdoulk,georgev\}@unipi.gr}}
\maketitle              

\begin{abstract}
Motivated by the global unrest related to the COVID-19 pandemic, we present a system prototype for ontology-based, integration of national data published from various countries. COVID-related data is published from different authorities, in different formats, at varying spatio-temporal granularity, and irregularly. Consequently, this hinders the joint data exploration and exploitation, which could lead scientists to acquire important insights, without having to deal with the cumbersome task of data acquisition and integration. Motivated by this shortcoming, we propose an approach for data acquisition, ontology-based data representation, and data transformation to RDF, which also enables interlinking with other publicly available data sources. Currently, data coming from the following European countries has been successfully integrated: Austria, Belgium, France, Germany, Greece, Italy, and Sweden. The knowledge base is automatically being updated, and it is available to the public through a SPARQL endpoint and a direct download link. Furthermore, we showcase how data integration enables spatio-temporal data analysis and knowledge discovery, by means of meaningful queries that would not be feasible to process otherwise.
\end{abstract}

	
\section{Introduction}
The COVID-19 virus outbreak is a major concern worldwide, as it affects both economically and socially every country in the world. Although the first reports had shown that the virus can be easily and sustainably transmitted between people, not all the countries took immediate measures to restrict the spreading of the virus. In Europe, different strategies and measures have been applied per country, with different outcomes, since the optimal way to respond was not clear. For example, Sweden relied at the ``herd immunity'' and no measures were taken against the virus for the first few weeks of the virus outbreak in the country, while Greece took immediate and restrictive measures when the first cases were detected~\cite{Georgiou2020.04.15.20066712}, isolating the population and cancelling international flights. 
An important factor contributing to the different strategies is the absence of high-quality, interlinked, globally accessible data, at various scales and levels of detail, to estimate the criticality of the condition at the very beginning of the virus spreading. Nevertheless, this lack of data as well as the need to effectively monitor the spread of the virus, while providing reports to the public, motivated most countries to collect data and provide daily reports on the virus spread and on the number of infected, recovered and deceased persons.


The wide variation of measures taken in connection to other data sources and their recorded results in each country can be analysed for future reference. Towards this goal, similar measures taken from different countries for their populations, need to be identified to safely evaluate their results. This task can be seen in general as a spatio-temporal analysis process, for pattern recognition, matching and prediction. Any such analysis task needs data from various countries, and for a large period of time. Additionally, a data set that combines reports from various countries and their regions can show the degree at which regions with similar characteristics have been affected and how they have affected other regions. As an example use case, correlation factors can be computed from the cases reported in adjacent regions of neighboring countries\footnote{A video generated from the compiled data set, that shows that there is possibly a correlation between adjacent regions, is available online at \url{http://83.212.169.101/datasets/covidOutbreak.html}}. 

Towards supporting complex data analysis and joint data exploration, it is necessary to aggregate COVID-19 data at different levels of administration regions, from different countries, and for a large period of time in an integrated data set. This means that data needs to be spatially and temporally aligned, and the spatio-temporal granularity at which data is recorded should be the same for all countries. 

In this work, we process a set of data sources providing national daily reports originating from different countries. 
We have developed an automated process, based on a system for RDF data transformation (RDF-Gen~\cite{WIMS2018}), that retrieves COVID-19 data from public sources and integrates them to RDF triples under a simple ontology. The ontology we use is built on top of well known ontologies such as OWL-Time, GeoSPARQL, and SIOC (Semantically-Interlinked Online Communities) Core Ontology. The resulting data set is linked to mainstream data providers such as wikiData and EU Open Data Portal. 

As such, the main contribution of this work is an approach for data integration of COVID-related data that enables complex data analysis queries that can extract useful knowledge and insights. 
Specifically, our work makes the following achievements:
\begin{itemize}
	\item We integrate the daily reports of COVID-19 from 7 central European countries, that have followed different strategies against the outbreak. We plan to expand the data set with more countries in the near future.
	\item We homogenize the COVID-19 reports of different countries under a single schema, while providing data at a specific level of spatial and temporal granularity, converting the spatial regions reported in the original sources to the corresponding regions using NUTS encoding for levels 1, 2, and 3.
	\item We show how data can be interlinked with the daily reports contextual data, such as population density per group age, total population and administration regions of countries (along with topological relations such as ``touches'' or ``contains'').
	\item We transforms the data into RDF using commonly used ontologies such as GeoSPARQL, RAMON, OWL-Time, SIOC (discussed briefly in Section \ref{ontology}), to enable linking additional data sources with third-party resources that are not already included in the compiled data set.
	\item We provide a SPARQL endpoint for data exploration through query answering, on the daily updated data set. The result set can be either rendered on a map or displayed as a table. The complete data set is also available online.
\end{itemize}

Section~\ref{relatedWork} briefly presents 
the data sources accessed for the compilation of this data set and the main issues tackled in the process. Then, in Section~\ref{ontology}, we present the COVID-19 ontology that we designed. Section~\ref{evaluation} briefly analyzes the data set of reported cases for the first 6 months of 2020 and for the selected countries. Finally, Section~\ref{conclusions} concludes the paper and sketches our future work.


\section{Data Sources and Data Acquisition}\label{relatedWork}

In this section, we provide a brief description of the data sources used in our work, along with the problems related to the published data. Then, we describe the process of data acquisition.

\subsection{Data Sources}

To the best of our knowledge, the publicly available data sets related to COVID-19 either report the total number of cases per country (NUTS\footnote{The regions in the retrieved data are converted to Nomenclature of Territorial Units for Statistics (NUTS) levels 1 to 3, since each data source uses a different reference system of regions.} level 0), or they report the number of cases for various administration regions within a specific country. Typical examples of the first category are the World Health Organization (WHO) Coronavirus Disease (COVID-19) Dashboard\footnote{\url{https://covid19.who.int/}}, the WorldOmeter\footnote{\url{https://www.worldometers.info/coronavirus/}} and the Johns Hopkins University dashboard\footnote{\url{https://coronavirus.jhu.edu/map.html}} and the related github project\footnote{\url{https://github.com/CSSEGISandData/COVID-19}}. However, these sources have been built mainly for reporting purposes rather than for doing any form of analysis. 
For example, it is not possible to retrieve the number of cases reported, at a specific day in the past and for a specific administration region of some country.

On the other hand, several countries provide daily reports about the numbers of cases per administration region. A summarizing page is available at Wikipedia\footnote{\url{https://en.wikipedia.org/wiki/COVID-19_pandemic}}, which also provides links to other Wikipedia pages maintained and updated by national health organizations of several countries. Nevertheless, it requires effort to extract the time series of reported cases for each country and their administration regions from the unstructured text.

In addition to the above, some organizations provide summaries of daily reports per region of a country, through github projects. Specifically, the daily reports of cases in Italy are available through the github project pcm-dpc/COVID-19 of ``Presidenza del Consiglio dei Ministri - Dipartimento della Protezione Civile''\footnote{\url{https://github.com/pcm-dpc/COVID-19}}. Similarly, the Greek daily reports can be accessed from the iMedd webpage\footnote{\url{https://www.imedd.org/new-covid-19-i-watch-the-spread-of-the-disease-in-greece-and-around-the-world/}} and the github project\footnote{\url{https://github.com/iMEdD-Lab}}. Since these data sources do not share a common schema (or even a common format), it is not always straightforward to relate the data provided, for obtaining a detailed view for several countries and investigate 
correlations or similar patterns of virus outbreak in different countries, at the same level of administrative regions.



\subsection{Data Acquistion}\label{dataSources}
The most rapid spreading of COVID-19 virus in Europe was detected in northern Italy. For this reason our data set was compiled starting from the Italian daily reports, and it is recursively expanded with reports of adjacent countries. We focus on adjacent countries that share common borders, since the transportation between regions is less likely to be affected by airport ``lockdowns''. Specifically, we access COVID-19 reports for the following countries:
\begin{itemize}
	\item Austria: data is retrieved and processed daily from the online service at \url{https://www.drawingdata.net/covmap/}
	\item Belgium: data is retrieved and processed daily from the dashboard at \url{https://epistat.wiv-isp.be/}
	\item France: data retrieved and processed from \url{https://www.data.gouv.fr/fr/datasets/donnees-hospitalieres-covid-19/}
	\item Germany: data is retrieved and processed from the online dashboard at \url{https://corona.rki.de}
	\item Greece: data retrieved and processed from the iMedD-Lab github project available online at \url{https://github.com/iMEdD-Lab/open-data/tree/master/COVID-19}
	\item Italy: data accessed and processed from the github project of ``Presidenza del Consiglio dei Ministri - Dipartimento della Protezione Civile'' available online at \url{https://github.com/pcm-dpc/COVID-19}
	\item Sweden: data retrieved and processed from the dashboard ``Sweden Coronavirus stats tracker'' at \url{https://visalist.io/emergency/coronavirus/sweden-country}
\end{itemize}

The following problems were addressed in the process of data integration from the above sources:
\begin{itemize}
	\item Different languages and encodings: Even in the same data source of a multi-language country (for example Belgium), it may occur that daily reports use different encodings (depending on the region and the language used in it). 
	This issue may result to incomplete data if not handled properly.
	\item Different types of sources and data formats (JSON, CSV, ESRI shapefiles): Data sources do not share a common format and schema. For example, data is provided in CSV tables, JSON files (e.g., GIS feature server responses), and ESRI shapefiles (e.g., EU Eurostat data). 
	\item Different encodings of spatial regions: Germany uses ``landkreis'' codes, Austria uses ``Gemeindekennziffer'' (GKZ) codes, Greek and Belgian sources use region names in all official languages. We tackled this issue using a key-value map between the region encoding used in the corresponding data source and NUTS level 3 codes. In the case of Belgium, which uses subsections of NUTS level 3 regions, the process aggregates the data into the corresponding NUTS level 3 region. This approach allows the spatial integration of data.
	\item Data is not temporally aligned and not updated in the same interval: Austrian reports are provided in a hourly interval, while Greek reports are not provided in a constant interval, and the rest of the data sources provide the data in a daily interval. 
	For this reason we retrieve the data of one day before current day, to guarantee that no further updates are expected for any reported day in our data set.
	\item Undefined spatial regions for reported cases (e.g., cases on a cruiser). We associate the reported cases with the nearest NUTS level 3 region (e.g., the port where the cruiser has docked).
	\item Different update strategies: Italian reports are provided in a separate file for each day in the corresponding github project. On the other hand, the github project for the Greek data overwrites previous reports, thus the timeline is only accessible through the git versioning process. Dashboard data sources (e.g., the data source used for the German reports) do not rely on specific files, but instead they are accessed as online endpoints.
\end{itemize}

In addition, missing records for some regions and days have been observed.


\section{The COVID-19 Ontology}\label{ontology}
The COVID-19 data set is integrated into an RDF graph using a new ontology that is based on the following imports:

\begin{itemize}
	\item SIOC\footnote{\url{https://www.w3.org/Submission/2007/02/namespaces.zip}}: The SIOC (Semantically-Interlinked Online Communities) Core Ontology provides the main concepts and properties required to describe information found on online communities such as social media, message boards, wikis, weblogs, etc.
	\item GeoSPARQL\footnote{\url{http://www.opengis.net/ont/geosparql}}: The OGC GeoSPARQL standard supports representing and querying geospatial data on the Semantic Web. GeoSPARQL defines a vocabulary for representing geospatial data in RDF, and it defines an extension to the SPARQL query language for processing geospatial data.
	\item OWL-Time\footnote{\url{https://raw.githubusercontent.com/w3c/sdw/gh-pages/time/rdf/time.ttl}}: OWL-Time is an ontology of temporal concepts, for describing the temporal properties of resources. The vocabulary provided expresses facts about relations among instants and intervals, as well as durations. Time positions and durations may be expressed using either the conventional (Gregorian) calendar and clock, or using another temporal reference system such as Unix-time, geologic time, or different calendars.
	\item RAMON geographic ontology\footnote{\url{https://ec.europa.eu/eurostat/ramon/ontologies/geographic.rdf}}: RAMON geographic ontology describes countries, NUTS, and Local Administrative Units (LAU) related concepts and properties.
\end{itemize}

\begin{figure}
	\centering
	\includegraphics[width=\linewidth]{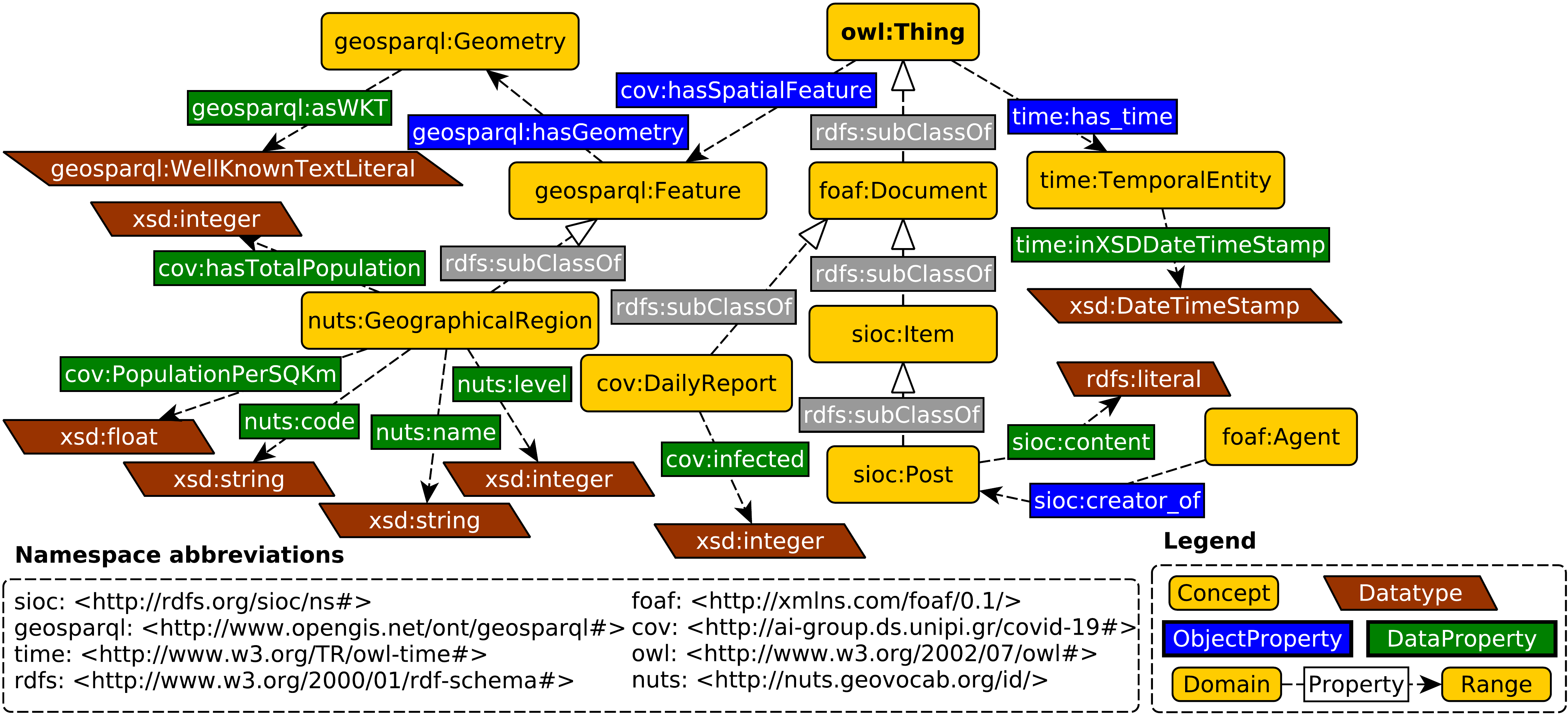}
	\caption{The concepts and properties combining SIOC, GeoSPARQL and OWL-Time ontologies for spatio-temporal-textual evaluation. Rounded rectangles in the graph represent concepts, edges illustrate properties and skewed parallelograms represent datatypes of literals.}
	\label{ont1}
\end{figure}

We combine the above ontologies using the added property \texttt{cov:hasSpatialFeature}, where \texttt{cov:} is the prefix for the namespace of COVID-19 ontology. The domain of the added property is \texttt{owl:Thing} and its range is \texttt{geosparql:Feature}. This means that any resource in our ontology can be related to a spatial feature, i.e.  locations of various natural or artificial boundaries or shapes to help visualize spatially-related data. Similarly, the domain of OWL-Time property \texttt{time:has\_time} is defined to be \texttt{owl:Thing}, thus any resource can have a temporal constituent. The part of the resulting schema that combines concepts and properties of the above ontologies, is illustrated in Figure \ref{ont1}. The rounded rectangles in the graph represent concepts, edges illustrate properties and skewed parallelograms represent datatypes.

\begin{figure}
	\centering
	\includegraphics[width=\linewidth]{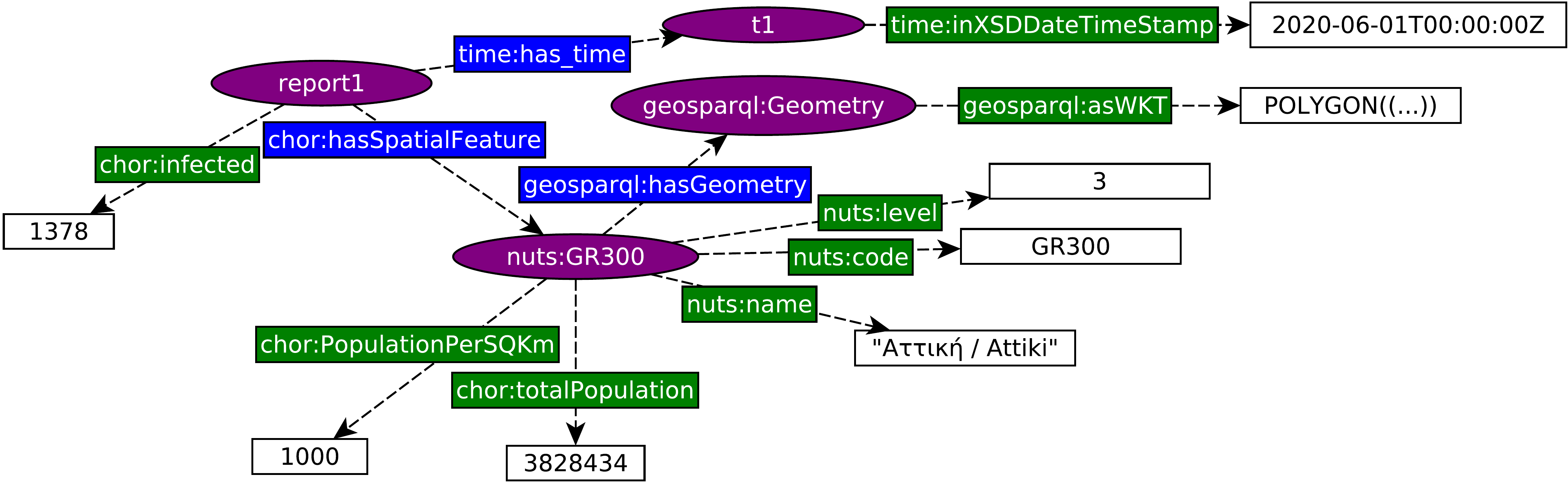}
	\caption{Example of RDF triples generated from COVID-19 data set.}
	\label{fig:example2}
\end{figure}

Regarding the association of geographical regions in NUTS geocode standard with COVID-19 reports, we define in the ontology that \texttt{nuts:GeographicalRegion} is a subclass of \texttt{geosparql:Feature}. We also introduce the concept \texttt{cov:DailyReport} for the set of daily reports provided in the COVID-19 data set. Each daily report is associated with geographical regions defined in RAMON ontology and GeoVocab\footnote{\url{http://nuts.geovocab.org/}} by the property \texttt{cov:hasSpatialFeature}. Finally, we introduce the data properties \texttt{cov:hasTotalPopulation}, \texttt{cov:PopulationPerSQKm}, \texttt{cov:populationGrpLE19}, \texttt{cov:populationGrp20\_39}, \texttt{cov:populationGrp40\_59}, \texttt{cov:populationGrpGE60}, for the total population, the population density and population per age group respectively. We add the property \texttt{cov:infected} for the reported infected cases of a region. The domain of these properties is defined to be the concept \texttt{nuts:GeographicalRegion}. Figure \ref{fig:example2} illustrates an example of triples representing a record from the COVID-19 data set under the defined schema. Please notice that the URI of the geographical region referred in the triples is in the RADON ontology namespace, which enables the exploitation of relations between geographical regions (e.g. aggregations of infected cases levels of administration regions), as specified in the ontology.

\begin{figure}
	\centering
	\includegraphics[width=\linewidth]{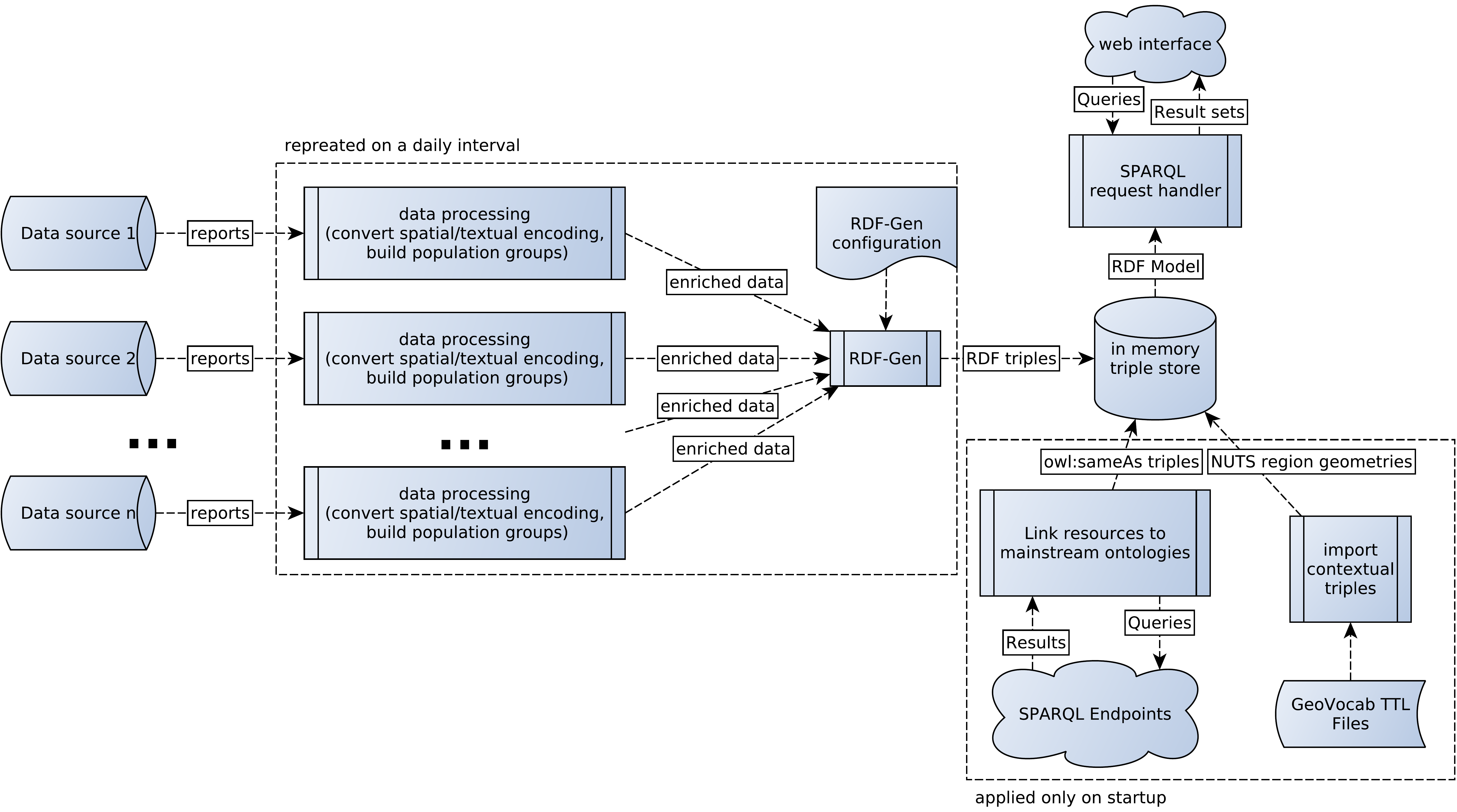}
	\caption{The overall workflow for retrieving, transforming updating and publishing data as RDF triples.}
	\label{fig:workflow}
\end{figure}

The transformation of data into RDF triples, is performed using RDF-Gen\cite{WIMS2018}. RDF-Gen transforms the data using a triples template, i.e. triples that allow the use of variables or predefined functions on any of the constituent parts (subject, predicate, object) of a triple. The process that connects to the public data, is initialized with contextual data related to the source, such as Administrative Regions and their geometries, population in each region and population groups per age. The enriched data are then provided to RDF-Gen, which generates the corresponding RDF triples. This process is automatic, repeated on a daily interval. The generated RDF triples are stored in a triple store, which allows the evaluation of federated SPARQL queries on the data set compiled. The triple store is initialized with GeoVocab TTL files that describe the geometries of NUTS regions and their topological relations. Figure \ref{fig:workflow} illustrates the overall workflow for the data set compilation.

Administrative regions in our data set are linked to EU Open Data Portal\footnote{\url{https://data.europa.eu/euodp/en/linked-data}} and wikiData\footnote{\url{https://www.wikidata.org/}}. The following query demonstrates the links to EU Open Data Portal:

\noindent\begin{adjustbox}{max width=\textwidth}
\begin{lstlisting}[backgroundcolor = \color{white},language=SPARQL,xleftmargin = 0.4cm,xrightmargin = 0.1cm,framexleftmargin = 1em, breaklines=false]
PREFIX : <http://ai-group.ds.unipi.gr/covid-19#> 
PREFIX owl:	<http://www.w3.org/2002/07/owl#>
SELECT * WHERE {
     ?r owl:sameAs ?u .
SERVICE <https://data.europa.eu/euodp/sparqlep> 
     {?u ?p ?o}
} LIMIT 10
\end{lstlisting}
\end{adjustbox}	

Similarly, the links to wikidata are demonstrated with the query:

\noindent\begin{adjustbox}{max width=\textwidth}
\begin{lstlisting}[backgroundcolor = \color{white},language=SPARQL,xleftmargin = 0.4cm,xrightmargin = 0.1cm,framexleftmargin = 1em, breaklines=false]
PREFIX : <http://ai-group.ds.unipi.gr/covid-19#> 
PREFIX owl:	<http://www.w3.org/2002/07/owl#>
PREFIX wd: <https://query.wikidata.org/>
SELECT * WHERE {
     ?r owl:sameAs ?u .
SERVICE wd:sparql
     {?u ?p ?o}
}
\end{lstlisting}
\end{adjustbox}

The linkage to wikidata graph, enables our data to be also combined with other graphs that are connected to wikidata, such as FactForge\footnote{\url{http://factforge.net/}}. For instance, the query:

\noindent\begin{adjustbox}{max width=\textwidth}
\begin{lstlisting}[backgroundcolor = \color{white},language=SPARQL,xleftmargin = 0.4cm,xrightmargin = 0.1cm,framexleftmargin = 1em, breaklines=false]
PREFIX : <http://ai-group.ds.unipi.gr/covid-19#>
PREFIX nuts: <http://nuts.geovocab.org/id/> 
PREFIX owl:	<http://www.w3.org/2002/07/owl#>
PREFIX ot: <http://ontology.ontotext.com/taxonomy/>
PREFIX ff: <http://factforge.net/repositories/>
SELECT ?s ?p ?o WHERE {
     ?r nuts:code "AT130" ; owl:sameAs ?u .
   SERVICE ff:ff-news {
      ?s ot:exactMatch ?u .
      ?s ?p ?o .
   }
}
\end{lstlisting}
\end{adjustbox}

\noindent will retrieve all triples related to the region in our data set, with NUTS code ``AT130'' (Vienna) from the FactForge graph.


\section{Evaluation}\label{evaluation}
The compiled data set is available online in two ways: a) through the SPARQL endpoint at \url{http://83.212.169.101/datasets/yasgui.html}, and b) as an archive at the download link \url{http://83.212.169.101/datasets/ttl.7z}. The SPARQL endpoint provides the user with spatial functions that can be used in the queries, and the option to render the results on a map (given that the results contain a spatial representation). In this case, the user can also define the variable that can be used to render the color of the geometries. If the variable used for coloring the geometries takes numerical values in the results, it automatically detects the minimum and maximum values and the color is set w.r.t. the value of the variable in the white-red spectrum. Otherwise, if the variable takes non-numerical values, the colors are assigned randomly.

We demonstrate the use of the compiled data set, on investigating whether the number of infections reported in adjacent regions of different countries have a linear correlation. The first query in this scenario, is to retrieve the adjacent regions that belong to different countries:\\
\begin{adjustbox}{max width=1.2\textwidth}
\begin{lstlisting}[backgroundcolor = \color{white},language=SPARQL,xleftmargin = 0.4cm,xrightmargin = 0.1cm,framexleftmargin = 1em]
PREFIX geosparql: <http://www.opengis.net/ont/geosparql#> 
PREFIX : <http://ai-group.ds.unipi.gr/covid-19#> 
PREFIX nuts: <http://nuts.geovocab.org/id/> 
PREFIX f: <java:SPARQL_functions.>

SELECT ?r1 ?r2 ?c1 ?wkt WHERE {
   ?r1 nuts:name ?name ; geosparql:hasGeometry/geosparql:asWKT ?wkt . ?c1 :hasPart ?r1 .
   ?r2 nuts:name ?name2 ; geosparql:hasGeometry/geosparql:asWKT ?wkt2 . ?c2 :hasPart ?r2 .
?c1 nuts:level "0" . ?c2 nuts:level "0" .
?r1 nuts:level ?l1 . ?r2 nuts:level ?l2 .
   FILTER(f:touches(?wkt,?wkt2)&&(?c1!=?c2) && 
      ((?l1="2")||(?l1="3")) && ((?l2="2")||(?l2="3")))
}
\end{lstlisting}
\end{adjustbox}

\begin{figure}
	\centering
	\includegraphics[width=0.8\linewidth]{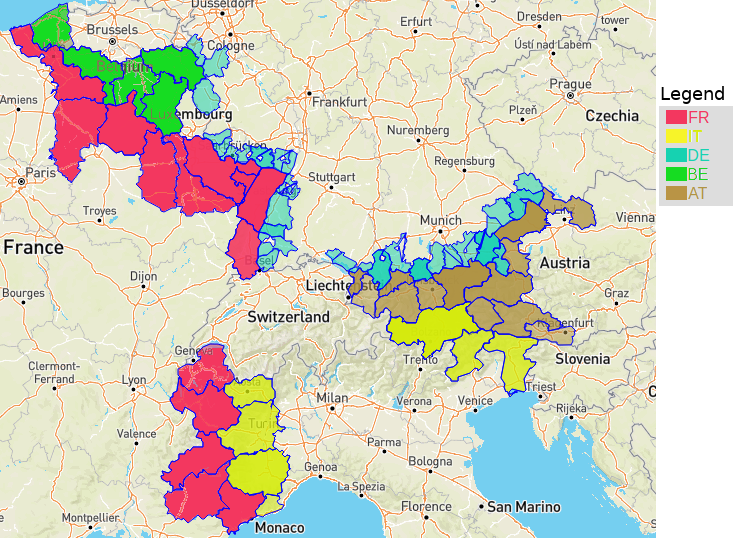}
	\caption{Adjacent regions of NUTS level 3, that belong to different countries.}
	\label{fig:map1}
\end{figure}

In this query we use the spatial function $touches(g1,g2)$, which returns true if only the geometries $g1,g2$ touch at their boundaries. Since the Belgian daily reports are assigned to regions at NUTS level 2, we include in the filter both levels ``2'' and ``3''. Figure \ref{fig:map1} illustrates the 67 pairs of regions returned in the result set of this query.

Including in the above query the triple patterns for the daily reports of infections in these regions, results to the following query:\\
\begin{adjustbox}{max width=\textwidth}
\begin{lstlisting}[backgroundcolor = \color{white},language=SPARQL,xleftmargin = 0.4cm,xrightmargin = 0.1cm,framexleftmargin = 1em ]
SELECT ?r1 ?r2 ?inf1 ?inf2 ?date WHERE {
   ?r1 nuts:name ?name ; geosparql:hasGeometry/geosparql:asWKT ?wkt . 
   ?c1 :hasPart ?r1 .
   ?r2 nuts:name ?name2 ; geosparql:hasGeometry/geosparql:asWKT ?wkt2 . 
   ?c2 :hasPart ?r2 .
?c1 nuts:level "0" . ?c2 nuts:level "0" .
?r1 nuts:level ?l1 . ?r2 nuts:level ?l2 .
?report1 :hasSpatialFeature ?r1 ; :infected ?inf1 ; 
       time:has_time/time:inXSDDateTimeStamp ?date.
   ?report2 :hasSpatialFeature ?r2 ; :infected ?inf2 ; 
       time:has_time/time:inXSDDateTimeStamp ?date.
   FILTER(f:touches(?wkt,?wkt2)&&(?c1!=?c2) && 
      ((?l1="2")||(?l1="3"))&& ((?l2="2")||(?l2="3")))
} ORDER BY ?date
\end{lstlisting}
\end{adjustbox}	

\begin{figure}
	\centering
	\includegraphics[width=0.9\linewidth]{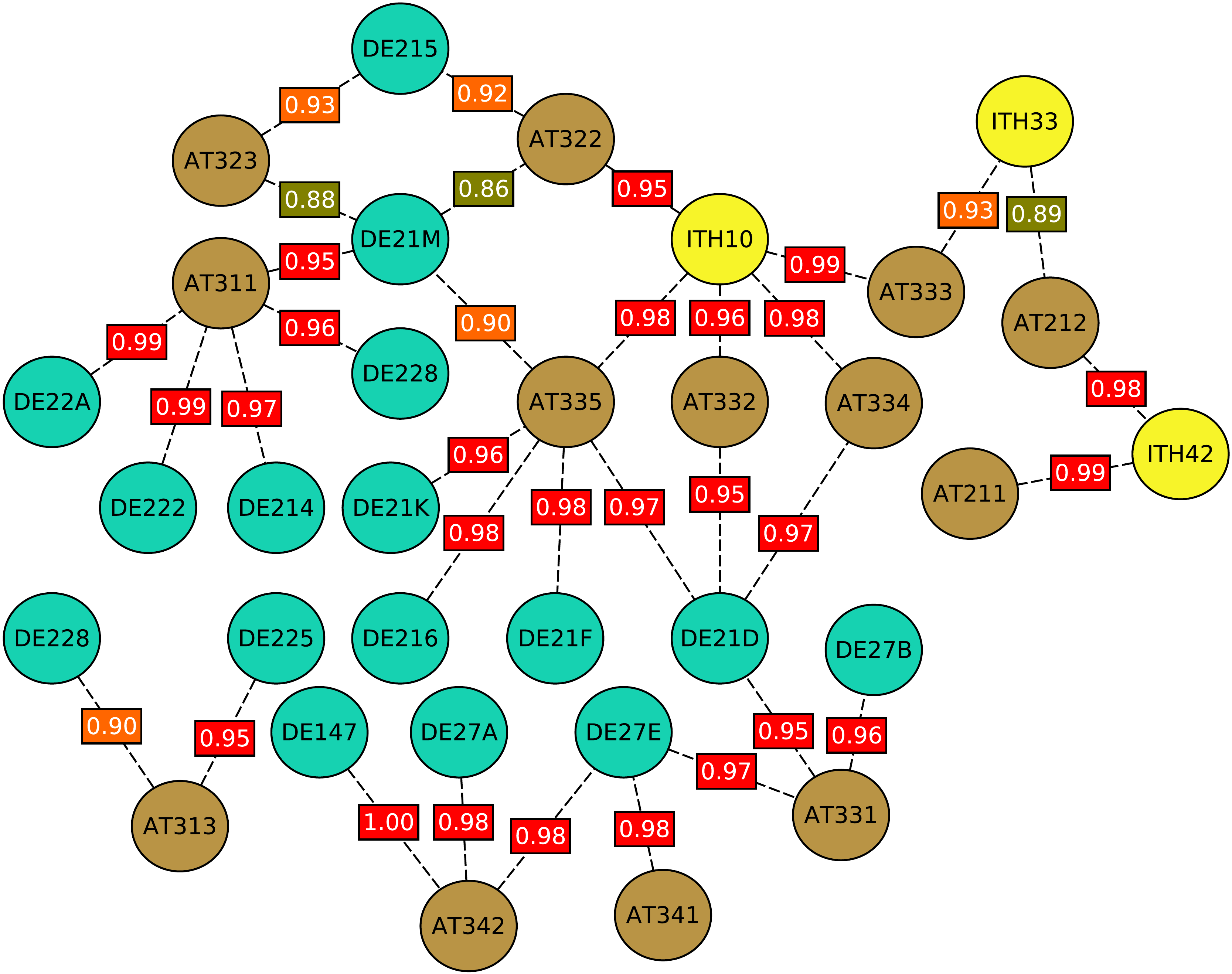}
	\caption{High correlations between adjacent regions. Nodes represent regions (NUTS level 3) and edges report the correlation value.}
	\label{fig:corr1}
\end{figure}

The results of the above query can be used to compute the Pearson correlation coefficient, on the number of cases reported on the adjacent regions. Interestingly, the computation of Pearson correlation, indicates three cases of correlation between regions, illustrated in the corresponding Figure \ref{fig:corr1}, and Tables \ref{tab:pearson2}, \ref{tab:pearson3}. Specifically, regions that seem to be highly affected from their neighbors (Pearson R value ranges between 0.86 and 1.00), are shown in Figure \ref{fig:corr1}. We observe that higher values of the coefficient correlation are between Austrian and German regions. The second case reported in Table \ref{tab:pearson2}, shows significantly lower correlation that varies between 0.25 and 0.46. The regions involved in this case are French and Belgian. Finally, Table \ref{tab:pearson3}, reports negative values of coefficient correlation, varying between -0.04 and -0.48. The negative values are found between French and Italian regions, and it can be a result of measures taken, when the reported numbers of infections were increased.

\begin{table}
  \begin{center}
    \caption{Regions with high coefficient correlation on the number of reported infections.}
    \label{tab:pearson2}
    \begin{tabular}{l|l|r} 
    \textbf{Region} & \textbf{Region} & \textbf{Pearson R}\\\hline
    FRF31 & BE34 & 0.46\\\hline
    FRF21 & BE35 & 0.45\\\hline
    BE34 & FRF32 & 0.42\\\hline
    BE32 & FRF21 & 0.39\\\hline
    FRE21 & BE32 & 0.38\\\hline
    FRE11 & BE32 & 0.32\\\hline
    FRE11 & BE25 & 0.28\\\hline
    BE34 & FRF21 & 0.25\\\hline
\end{tabular}
  \end{center}
\end{table}

\begin{table}[h!]
  \begin{center}
    \caption{Regions with high coefficient correlation on the number of reported infections.}
    \label{tab:pearson3}
    \begin{tabular}{l|l|r} 
    \textbf{Region} & \textbf{Region} & \textbf{Pearson R}\\\hline
FRF33 & DEB3K & -0.04\\\hline
FRF11 & DEB3K & -0.08\\\hline
DE133 & FRF11 & -0.10\\\hline
FRK28 & ITC20 & -0.11\\\hline
ITC31 & FRL03 & -0.11\\\hline
FRF11 & DEB3H & -0.11\\\hline
FRL03 & ITC16 & -0.14\\\hline
BE33 & DEA2D & -0.15\\\hline
FRF11 & DE124 & -0.18\\\hline
FRF33 & DEC05 & -0.19\\\hline
FRF33 & DEC01 & -0.21\\\hline
ITC16 & FRL01 & -0.22\\\hline
FRF33 & DEC02 & -0.22\\\hline
FRK27 & ITC20 & -0.23\\\hline
FRF12 & DE133 & -0.24\\\hline
FRF33 & DEC04 & -0.25\\\hline
BE33 & DEB23 & -0.26\\\hline
FRF11 & DEB3E & -0.33\\\hline
DE134 & FRF11 & -0.34\\\hline
FRF12 & DE132 & -0.35\\\hline
BE33 & DEA28 & -0.37\\\hline
DE139 & FRF12 & -0.38\\\hline
FRL02 & ITC16 & -0.43\\\hline
FRL02 & ITC11 & -0.46\\\hline
ITC11 & FRK27 & -0.48\\\hline
\end{tabular}
  \end{center}
\end{table}


%




\section{Conclusions}\label{conclusions}
This work presents a prototype for generating and updating an Open Data set about COVID-19 confirmed cases that enables cross-country analysis at different levels of granularity. 
The compiled data set is transformed into RDF triples to populate an ontology built on top of well-known ontologies, and resources are linked to external Open Data repositories. We plan to expand our data set with more countries and link the data with more portals that provide information about social events, news feeds and human activities that possibly affect the spreading of the virus.

\bibliographystyle{abbrv}
\bibliography{references}

\end{document}